\documentclass{natureprintstyle}
\usepackage{epsfig,caption}
\usepackage{color}
\usepackage{bm}
\usepackage{graphicx}
\usepackage{longtable}
\usepackage{amssymb}
\usepackage{rotating}

\newcommand{\apj}{Astrophys. J.}

\newcommand{\apjs}{Astrophys. J. Supp.}
\newcommand{\araa}{Annu. Rev. Astron. Astrophys.}
\newcommand{\mnras}{Mon. Not. R. Astron. Soc.}
\newcommand{\apjl}{Astrophys. J. Let.}
\newcommand{\aap}{Astron. Astrophys.}

\newcommand{\nat}{Nature}
\newcommand{\na}{New Astron. Rev.}

\usepackage{hyperref}

\title{The Formation of Submillimetre-Bright Galaxies from Gas Infall
  over a Billion Years}

\author{Desika Narayanan$^{1,13}$, Matthew Turk$^2$, Robert Feldmann$^{3,12}$,  Thomas Robitaille$^4$, Philip Hopkins$^5$, Robert Thompson$^{6,7}$,  Christopher Hayward$^{5,8}$, David Ball$^{7,9}$, Claude-Andr\'e Faucher-Gigu\`ere$^{10}$, Du\v{s}an Kere\v{s}$^{11}$}

\begin{document}

\maketitle

\let\thefootnote\relax\footnote{

%\begin{affiliations}
%\item Haverford College, 370 W Lancaster Ave, Haverford, PA 19041, USA
%\end{affiliations}
%}

\begin{affiliations}
\item Haverford College, 370 W Lancaster Ave, Haverford, PA 19041, USA

\item National Center for Supercomputing Applications, University of
  Illinois, 1205 W. Clark Street, Urbana-Champaign, IL, 61820, USA
  
\item Department of Astronomy and Theoretical Astrophysics Center, University of California, Berkeley, CA 94720 USA
  
\item Max Planck Institute for Astronomy, Konigstuhl 17, D-69117, Heidelberg, Germany

\item TAPIR, California Institute of Technology, MC 350-17, Pasadena, CA 91125, USA

\item University of the Western Cape, Bellville, Cape Town, South Africa, 7535

\item Steward Observatory, University of Arizona, 933 N Cherry Ave, Tucson, AZ 85721

\item Harvard-Smithsonian Center for Astrophysics, 60 Garden St. Cambridge, MA 02138 

\item Whitman College, 345 Boyer Ave, Walla Walla, WA, 99362, USA

\item CIERA, Northwestern University, 2145 Sheridan Road, Evanston, IL, 60208

\item CASS, University of California, San Diego, 9500 Gilman Drive, La Jolla, CA 92093

\item Hubble Fellow

\item Corresponding Authour
\end{affiliations}
}

\vspace{-3.5mm}
\begin{abstract}

  Submillimetre-luminous galaxies at high-redshift are the most luminous,
  heavily star-forming galaxies in the Universe\cite{casey14a}, and
  are characterised by prodigious emission in the far-infrared at 850
  microns ($S_{\rm 850} \geq 5$ mJy).  They reside in halos $\sim
  10^{13} M_\odot$\cite{hickox12a}, have low gas fractions compared to
  main sequence disks at a comparable redshift\cite{geach11a}, trace
  complex environments\cite{fu13a,daddi09a}, and are not easily
  observable at optical wavelengths\cite{swinbank04a}.  Their physical
  origin remains unclear.  Simulations have been able to form galaxies
  with the requisite luminosities, but have otherwise been unable to
  simultaneously match the stellar masses, star formation rates, gas
  fractions and
  environments\cite{baugh05a,hayward13a,shimizu12a,dave10a}.  Here we
  report a cosmological hydrodynamic galaxy formation simulation that
  is able to form a submillimetre galaxy which simultaneously
  satisfies the broad range of observed physical constraints.  We find
  that groups of galaxies residing in massive dark matter halos have
  rising star formation histories that peak at collective rates $\sim
  500-10^3 M_\odot {\rm yr}^{-1}$ at $z = 2-3$, by which time the
  interstellar medium is sufficiently enriched with metals that the
  region may be observed as a submillimetre-selected system. The
  intense star formation rates are fueled in part by a reservoir gas
  supply enabled by stellar feedback at earlier times, not through
  major mergers. With a duty cycle of nearly a gigayear, our
  simulations show that the submillimetre-luminous phase of high-z
  galaxies is a drawn-out one that is associated with significant mass
  buildup in early Universe proto-clusters, and that many
  submillimetre-luminous galaxies are actually composed of numerous
  unresolved components (for which there is some observational
  evidence\cite{hodge13a}).

\end{abstract}

 We have conducted our cosmological hydrodynamic zoom galaxy formation
 simulations utilising the new hydrodynamic code {\sc
   gizmo}\cite{hopkins14a}, and include a model for the impact of
 radiative and thermal pressure from stars on the multiphase
 interstellar medium.  This feedback both regulates the star formation
 rate, and shapes the structure in the interstellar medium.  Informed
 by clustering measurements of observed SMGs\cite{hickox12a}, we focus
 on a massive ($M_{\rm DM} \sim 10^{13}$ M$_\odot$ at $z=2$) halo with
 baryonic particle mass $M_{\rm bary} \sim 10^5 M_\odot$ as the host
 of our ``main galaxy'', and run the simulation to $z=2$.  The only
 condition of the tracked galaxy pre-selected to match the physical
 properties of observed SMGs is the chosen halo mass.  We combine this
 with a new dust radiation transport package, {\sc powderday}, that
 simulates the traverse of stellar photons through the dusty ISM of
 the galaxy, allowing us to robustly translate our hydrodynamic
 simulation into observable measures.  We simulate the radiative
 transfer from a 200 kpc region around the main galaxy.  This
 simulation represents the first cosmological model of a galaxy this
 massive to be explicitly coupled with dust radiative transfer
 calculations.  The details of both codes and the simulation setup are
 fully described in the Methods section.

We define two distinct regions in the simulations.  The
``submillimetre emission region'' is the 200 kpc region surrounding
the central galaxy in the halo of interest.  This is the region where
all of the modeled $850 \mu$m emission comes from, and is what relates
most directly to observations.  The ``submillimetre galaxy'' refers to
the central galaxy in the halo.  Physical quantities from the
submillimetre galaxy are most applicable to high-resolution
observations, as well as placing these models in the context of other
theoretical galaxy formation models.  As we will show, the
submillimetre emission from the region is generally dominated by the
central submillimetre galaxy, though the contribution from lower mass
galaxies is often non-negligible.

We track the submillimetre properties of the galaxies within the
region from $z\sim 6$.  The star formation rates of galaxies in the
region rise from this redshift toward later times $z\approx 2$, owing
to accretion of gas from the intergalactic medium (Figure 1).  As
stars form, stellar feedback-driven galactic winds generate outflows
and fountains allowing recycled gas to be available for star formation
at later times (ED Fig 1).  This
phenomena shapes a star formation history that is still rising at $z
\sim 2$, in contrast to galaxy formation models with more traditional
implementations of subresolution feedback, which peak at $z \sim 3-6$
for galaxies of this mass\cite{dave12a,hopkins14b,feldmann14a}.
Mergers and global instabilities drive short-term variability in the
global SFR, while outflows and infall driven by the feedback model
can impact features in the star formation history with a somewhat
cyclical 'saw-tooth' pattern.

At its earliest stages ($z \sim 4-6$), the integrated SFR from the
galaxies in the region varies from $\sim 100-300 \ M_\odot$ yr$^{-1}$,
with a significant stellar mass ($0.5-1 \times 10^{11} M_\odot$) in
place, comparable to some high-redshift
detections\cite{finkelstein13a}.  Feedback from massive stars enrich
the interstellar medium with metals, and the dust content
simultaneously rises.  By $z \approx 3$, the combination of gas
accumulation with substantial metal enrichment drives a factor of
$\sim 50$ increase in the dust mass, with masses approaching $\sim 1
\times 10^{9} M_\odot$.  Radiation from the delayed peak in the star
formation rate interacting with this substantive dust reservoir drives
the observed $850 \mu$m flux density to detectable values of $>5$ mJy.
The galaxies associated with the main halo enter a long-lived
submillimetre-luminous phase, with a duty cycle of $\sim 0.75$ Gyr.
While our main model is only run to $z=2$ owing to computational
restrictions for models of this resolution, tests with
lower-resolution models reveal that at later times ($z < 1.5$), a
declining star formation rate due to inefficient accretion as well as
exhausted gas supply drives a drop in the submillimetre flux density
(for more details, see the Methods section).  The star formation
history of galaxies residing in ($z=2$) $M_{\rm DM} \approx 10^{13}$
M$_\odot$ halos as controlled by the underlying stellar feedback
provides a physical explanation for the peak in the observed SMG
redshift distribution at $z=2-3$\cite{weiss13a}.

During the submillimetre-luminous phase, the emitting region is almost
always occupied by multiple detectable galaxies.  In Figure 2, we
present gas surface density projections of six arbitrarily chosen
snapshots during the evolution of the submillimetre-luminous phase
($z=2-3$).  The panels are $250$ kpc on a side; for reference, the
full width at half maximum (FWHM) of the Submillimetre Common-user
Bolometer Array (SCUBA) on the James Clerk Maxwell Telescope (JCMT),
the first instrument to detect SMGs, is $\sim 125$ kpc at $z\approx
2$.  Multiple clumps of gas falling into the central are nearly always
present.  The observed flux density from the region is typically
dominated by the central, with (on average) $\sim 30\%$ arising from
emission from subhalos (ED Fig 2).  The submillimetre flux density of
the central galaxy rises dramatically between $z\sim 2-3$, reaching a
peak value of $\sim 20$ mJy.  Owing to contributions from subhalos
surrounding the central, the flux from the overall $200$ kpc region
can exceed this, peaking at $\sim 30$ mJy.  Similarly extreme systems
have recently been detected with the Herschel Space Observatory and
South Pole Telescope\cite{ivison13a,fu13a,hezaveh13a}.

While the central is being bombarded by subhalos over a range of mass
ratios during the submillimetre-luminous phase, major galaxy mergers
akin to local prototypical analogues such as Arp 220 or NGC 6240 do
not drive the onset of the long-lived submillimetre-luminous phase in
the central galaxy.  In Figure 1, we highlight when the galaxy
undergoes a major merger with mass ratio $\geq 1:3$.  While major
mergers are common at early times (and indeed drive some short-lived
bursts in star formation), the bulk of the submillimetre-luminous
phase at later times ($z\approx 2-3$) occurs nearly a gigayear after
the last major merger.  The ratio of the SFR to its integral over
cosmic time (the specific SFR) of the overall emitting region is
generally on the main sequence of galaxy formation at $z \sim 2$
(defined as the main locus of points on the SFR-$M_*$ relation),
though the central can have values comparable both to main sequence
galaxies between $z \sim 2-3$, as well as outliers.  One consequence
of a model in which SMGs typically lie on the main sequence of star
formation is that the gas surface densities show a broad range, from
$\sim 100-10^{4} M_\odot {\rm pc}^{-2}$
(ED Fig 3), as well as diverse gas
spatial extents (Figure~\ref{figure:extent}).  This is manifested
observationally in the broad swath occupied by SMGs in the
Kennicutt-Schmidt star formation relation\cite{casey14a}.  The gaseous
spatial extent and surface densities is to be contrasted, however,
with local merger-driven ultraluminous infrared galaxies (ULIRGs),
which exhibit typical full width at half maximum radii of $\sim
100-500$ pc\cite{downes98a}.  Idealised galaxy merger simulations with
initial conditions designed to form SMGs further underscore this
contrast, as they also result in compact morphologies during final
coalescence, and can be inefficient producers of submillimetre
radiation owing to increased dust temperatures\cite{hayward13a}.

The central submillimetre galaxy is amongst the most massive and
highly star-forming of galaxies at this epoch.  The stellar masses are
diverse, and range from $\sim 1-5 \times 10^{11}$ M$_\odot$,
comparable to recent measurements of this
population\cite{michalowski12a}, as well as constraints from abundance
matching techniques\cite{behroozi12a}.  The molecular gas fractions of
the central galaxy ($f_{\rm gas} \equiv M_{\rm H2}/(M_{\rm H2}+M_*)$)
decline with stellar mass, and range from $\sim 40\%$ at lower stellar
masses to $\lesssim 10\%$ at the highest masses.  This is in agreement
with observations\cite{tacconi13a}, though is dependent on the
conversion from carbon monoxide ($^{12}$CO) luminosity to H$_2$ gas
mass.  We note that these predictions are quantitatively different
from those produced by previous cosmological efforts in this field,
with some predicted gas fractions exceeding $f_{\rm gas}=
0.75$\cite{baugh05a,shimizu12a} and median stellar masses as low as
$\sim 10^{10}$ M$_\odot$\cite{baugh05a}.  We present the plots
highlighting the gas fractions, and calculated SEDs of our model SMG
in the context of observations in ED Figs 4-5.  The gas distributions
within the central galaxy, which range from $\sim 1-8$ kpc, compare
well with recent observed dust maps with the Atacama Large Millimetre
Array\cite{simpson15a}.

The stellar masses, gas fractions and duty cycles are in agreement
with previous lower-resolution cosmological efforts\cite{dave10a},
though the predicted SFR and luminosity from this model are
substantially larger.  The star formation rate of the group of
galaxies in the region peaks at $\sim 1500 M_\odot$ yr$^{-1}$.  Importantly, 
up to half of the total infrared luminosity can come from older stars
with ages $t_{\rm age} > 0.1 $Gyr.  Utilising standard
conversions\cite{kennicutt12a}, the estimated star formation rate from
the integrated infrared SED ($3-1100 \mu$m) can exceed $\sim 3000
M_\odot$ yr$^{-1}$ (ED Fig 6), and hence
infrared-based star formation rate derivations of dusty galaxies at
high-$z$ may over-estimate the true SFR by a factor $\sim 2$.  Indeed,
the contribution of satellite galaxies to the global SFR, alongside
the contribution of old stars to the infrared luminosity may relieve
some tensions between the inferred star formation rates from
submillimetre galaxies and massive galaxies modeled in cosmological
hydrodynamic simulations\cite{dave10a}.

The end-product of the central submillimetre galaxy at $z \sim 2$ is a
galaxy with a stellar mass of $\sim 4-5 \times 10^{11} M_\odot$,
distributed over a similarly compact region of $\sim 1-5$ kpc as the
gas (Figure~\ref{figure:extent}).  This is similar in extent and mass to the $z \sim 2$
compact quiescent galaxy population, an observed population with mean
half-light radius of $R_e \approx 1.5$ kpc, stellar mass $M_* >
10^{11} M_\odot$, and ages $t_{\rm age} \sim 0.5-1$
Gyr\cite{vandokkum08b}, suggesting a plausible connection between the
galaxy populations.  Indeed, a calculation of the stellar velocity
dispersion along three orthogonal sightlines of the central during the
submillimetre-luminous phase results in $\sigma_* \approx 600-700$ km
s$^{-1}$, comparable to measurements of high-$z$ compact
quiescents.  A large sample of simulated SMGs will
allow for a robust analysis of expected abundances of SMGs and compact
quiescents at $z \sim 2$.

Our picture for SMG formation suggests that they are not transient
events, but rather natural long-lived phases in the evolution of
massive halos.  The $\sim 0.75$ Gyr duty cycle combined with the
comoving abundance\cite{murray13a} of dark matter halos of this mass
results in an expected abundance of our model SMGs of $\sim 1.5 \times
10^{-5} h^3$ Mpc$^{-3}$, comparable to the $\sim 10^{-5} h^3$
Mpc$^{-3}$ observed for SMGs\cite{chapman05a}.  While modeling the
full number counts involves convolving the typical duty cycle as a
function of halo mass with halo mass functions over a range of
redshifts, the approximate abundances implied by this model are
encouraging.

This model suggests that galaxies that form in halos of mass $M_{\rm
  DM} \approx 10^{14}$ at $z=0$ will represent typical SMGs near the
peak of their redshift distribution.  Lower mass models show that they
do not achieve the requisite star formation rate and metal enrichment
to generate submillimetre-luminous galaxies (see Methods section).
More extreme SMGs being detected between
$z=5-6$\cite{riechers13a,vieira13a} may form in even more massive
(and rare) halos than those considered here.

%\bibliography{/Users/desika/Dropbox/paper/full_refs}

\begin{addendum}
\item [Acknowledgements] The authours thank
  Micha{\l}~J.~Micha{\l}owski for providing observational
  data. Partial support for DN was provided by NSF AST-1009452,
  AST-1442650, NASA HST AR-13906.001, and a Cottrell College Science
  Award.  PFH, CCH, MT and RT were funded by the Gordon and Betty
  Moore Foundation (GBMF4561 and Grant \#776).  PFH acknowledges the
  Alfred P.\ Sloan Foundation for support.  CAFG was supported by NASA
  awards PF3-140106, NNX15AB22G, and NSF AST-1412836.  DK was
  supported by NSF AST-1412153.  RF was supported by NASA
  HF-51304.01-A .  The simulations here were run on Stampede at TACC
  through NSF XSEDE allocations \#TG- AST120025, TG-AST130039, and
  TG-AST140023, NASA Pleiades, and the Haverford College cluster.

\item[Author Contributions] D.N. wrote the text, and led the radiative
  transfer simulations and analysis.  D.N., M.T., T.R. and R.T. wrote
  the {\sc Powderday} software. R.T., C.C.H. and D.B. contributed to
  simulation analysis, and R.F., P.H., C-A.F-G and D.K. performed the
  cosmological simulations

\item[Author  Information]  Reprints  and permissions  information  is
  available  at   www.nature.com/reprints.  The  authors   declare  no
  competing financial interests. Readers are welcome to comment on the
  online  version  of  the  paper.  Correspondence  and  requests  for
  materials should be addressed to D.N. (dnarayan@haverford.edu).

\end{addendum}

%\begin{thebibliography}{10}
%\expandafter\ifx\csname url\endcsname\relax
%  \def\url#1{\texttt{#1}}\fi
%\expandafter\ifx\csname urlprefix\endcsname\relax\def\urlprefix{URL }\fi
%\providecommand{\bibinfo}[2]{#2}
%\providecommand{\eprint}[2][]{\url{#2}}
\newpage
\begin{figure*}
\centerline{
\includegraphics[width=0.9\textwidth]{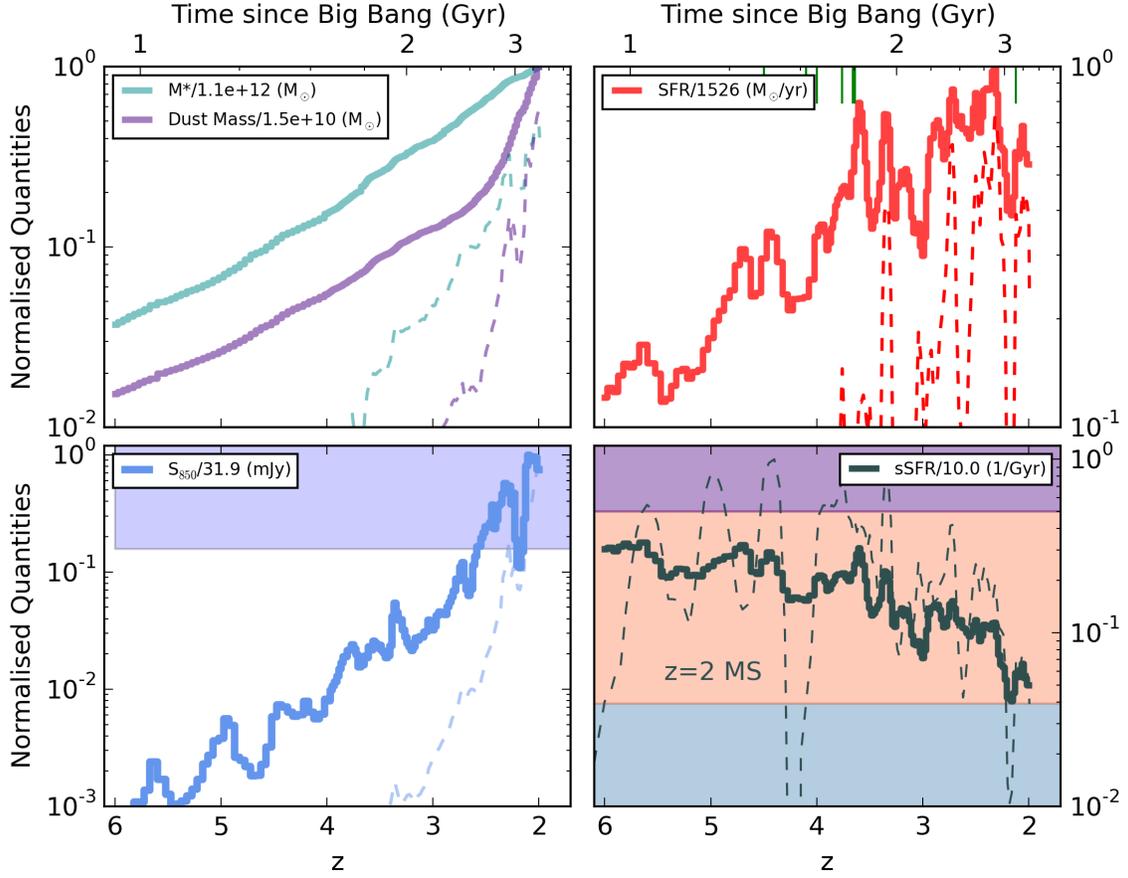}
}
\vspace{-4mm}
\caption{\textbf{Evolution of physical and observable properties of
    submillimetre emission region and central galaxy.}  The $200$ kpc
  submillimetre emission region are shown with thick solid lines,
  while the central galaxy's properties are given by thin dashed lines
  in each panel. Stellar and dust mass are in the top left, SFR at the
  top right; predicted observed 850 $\mu$m flux density at the bottom
  left; and specific SFR ($M_*/$SFR) at the bottom right.  The SFR is
  averaged on 50 Myr timescales, and includes a correctional factor
  $0.7$ for mass loss.  Locations of major galaxy mergers ($>1:3$) are
  noted by green vertical ticks on the top axis of the top right
  panel.  The blue shaded region in the 850 $\mu$m curve shows when
  the galaxy would be detectable as an SMG with SCUBA ($S_{\rm 850} >
  5$ mJy).  The yellow and purple shaded regions in the bottom right
  show the rough ranges for the $z=2$ Main Sequence and Starburst
  regime. The grey region denotes below the Main Sequence. }
\vspace{-4mm}
\end{figure*}

\begin{figure*}
\centerline{
\includegraphics[width=0.9\textwidth]{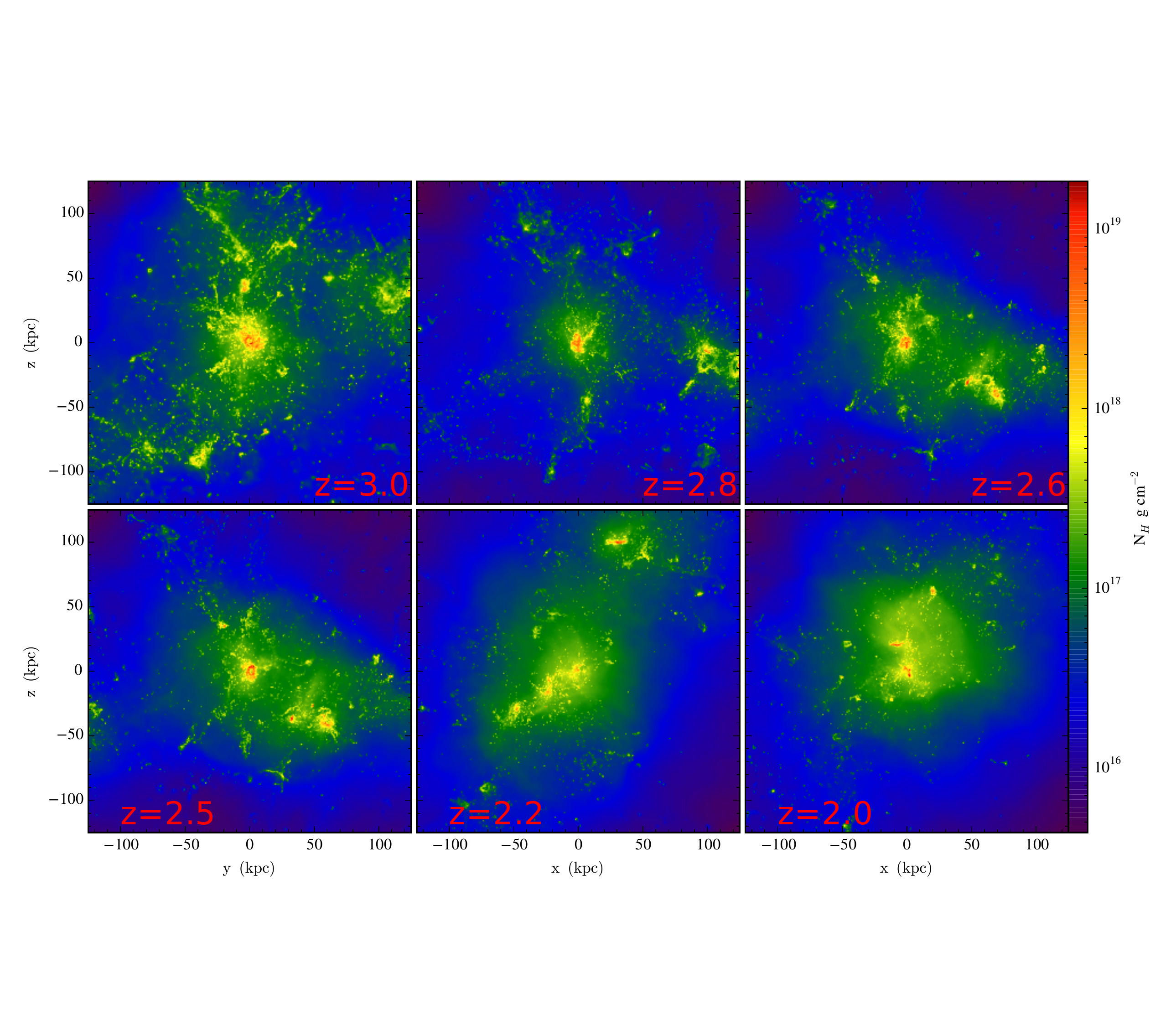}
}
\vspace{-4mm}
\caption{\textbf{Surface density projection maps of $250$ kpc region
    around central submillimetre galaxy between redshifts $z \approx
    2-3$.}  The submillimetre-emission region probed in surveys
  typically encompasses a central galaxy in a massive halo that is
  undergoing a protracted bombardment phase by numerous sub-halos.
  Some of the brightest SMGs arise from numerous galaxies within the
  beam in a rich environment (bottom right panel).}
\vspace{-4mm}
\end{figure*}

\begin{figure*}
\centerline{
\includegraphics[width=0.9\textwidth]{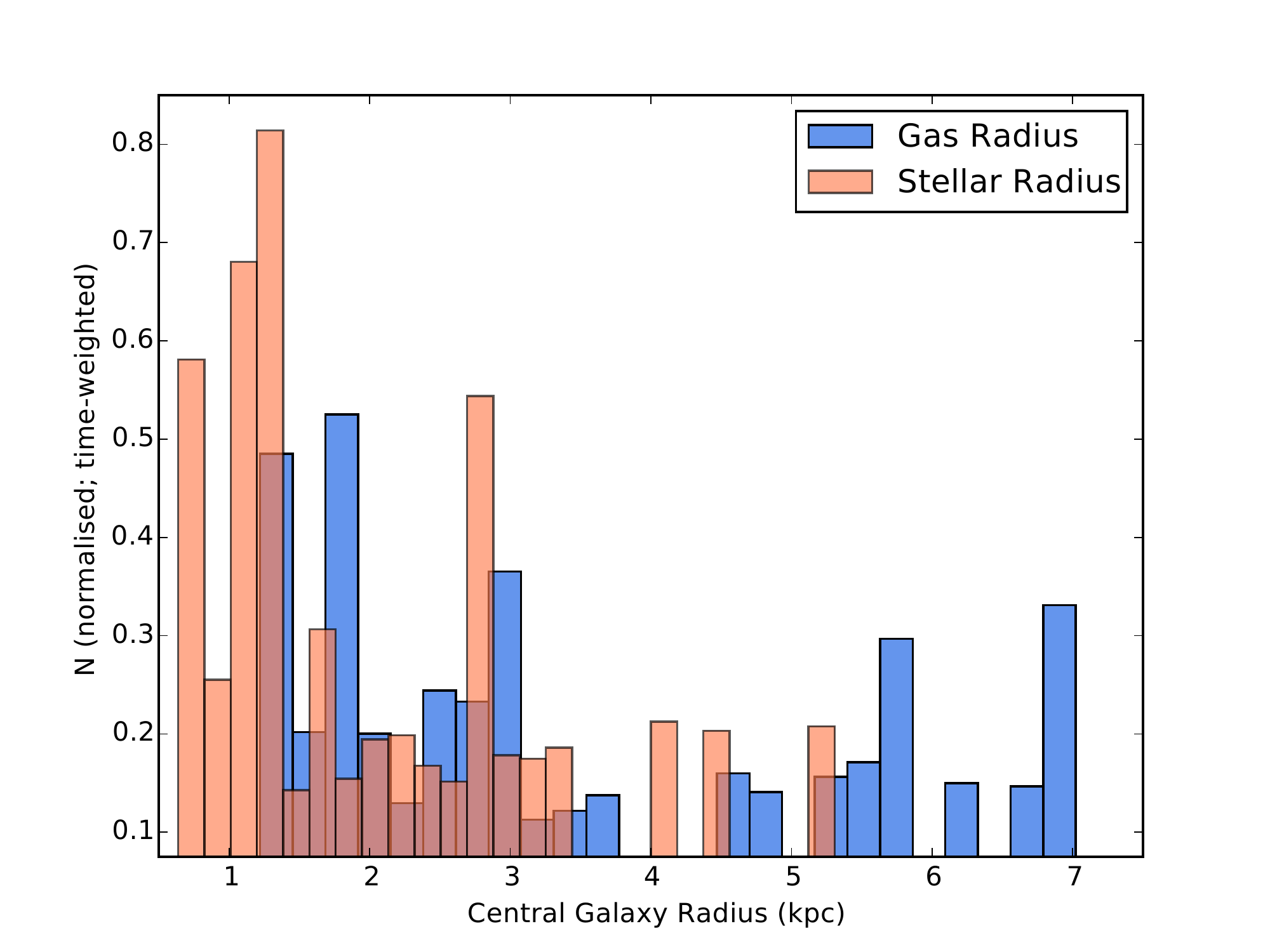}
}
\vspace{-4mm}
\caption{\textbf{Gas and stellar radius distribution for central
    submillimetre galaxy.}  The orange histogram denotes the half mass
  radius of the stars, while the blue shows the gas.  The galaxy gas
  is more distributed in the central than the (sub-kiloparsec) extent
  expected from major mergers, though still sufficiently compact that
  it will remain unresolved even with $\sim$ arcsecond resolution.\label{figure:extent}}
\vspace{-4mm}
\end{figure*}

\clearpage

\setcounter{page}{1}
\setcounter{figure}{0}
\setcounter{table}{0}
\renewcommand{\thefigure}{S\arabic{figure}}
\renewcommand{\thetable}{S\arabic{table}}

%\begin{document}
%\maketitle

\begin{center}
{\bf \Large \uppercase{Methods} }
\end{center}

\section{Cosmological Hydrodynamic Zoom Simulations}
We utilise a newly-developed version of TreeSPH that employs a
pressure-entropy formulation of smoothed particle hydrodynamics
(SPH)\cite{hopkins13d} that obviates many of the potential
discrepancies noted between grid-based codes, traditional SPH codes,
and moving-mesh algorithms\cite{agertz07a,sijacki12a,hayward13c}.  In
particular, we employ the hydrodynamic code {\sc
gizmo}\cite{hopkins14a} in P-SPH mode which conserves momentum,
energy, angular momentum and entropy, and includes newly developed
algorithms to treat the artificial viscosity, entropy diffusion and
time-stepping\cite{hopkins13d,fauchergiguere14a}.  The gravity solver
is a modified version of the {\sc gadget-3} solver\cite{springel05a},
and an updated softening kernel to better represent the potential of
the SPH smoothing kernel is included\cite{barnes12a}.

The simulations are fully cosmological zoom-in calculations of the
evolution of individual galaxies.  A 144 Mpc$^3$ cosmological volume
was simulated at low resolution down to redshift $z=0$ with dark
matter only.  The halo of interest was identified, and re-simulated at
much higher resolution with baryons included.  The initial conditions
were generated with the {\sc music} code\cite{hahn11a}.  We simulate
four zoom galaxies -- one is our main galaxy, and the other three are
at varying resolutions and masses for the purposes of testing.  The
main galaxy of interest to this study resides in a dark matter halo
mass of $M_{\rm DM} = 3 \times 10^{13}$ M$_\odot$ at $z=2$.  The
initial baryonic particle masses in the high-resolution region were
$2.7 \times 10^5 M_\odot$, and the minimum baryonic/stars/dark matter
force softening lengths were $9/21/142$ proper pc at $z = 2$.  The
physical properties of all of the modeled galaxies are presented in ED
Table 1 in the Extended Data.

The baryonic physics implemented into {\sc gizmo} are developed based
on extensive tests studying idealised simulations of both isolated
disks and galaxy
mergers\cite{hopkins11b,hopkins12a,hopkins13a,hopkins13b,hopkins13c,narayanan13a}.
The gas cools utilising an updated cooling curve to
standard\cite{katz96a} implementations in SPH codes which includes
both atomic and molecular line emission\cite{ferland13a}. The modeled
interstellar medium is multiphase.  The neutral ISM is broken into an
atomic and molecular component following algorithms that scale the
molecular fraction with column density and gas phase
metallicity\cite{krumholz08,krumholz11b}.  Star formation occurs in
molecular gas above a threshold density (here, this is set to $n_{\rm
thresh} = 10$ cm$^{-3}$).  Star formation is further restricted to gas
that is locally self-gravitating, where:
\begin{equation}
\alpha \equiv \beta^{\prime}\,\frac{|\nabla\cdot {\bf v}|^{2} + |\nabla\times {\bf v}|^{2}}{G\,\rho} < 1
\end{equation}
  This follows from studies\cite{hopkins13e} that show that the
predicted spatial distribution of star formation in galaxies is more
realistic when utilising a gas self-gravitating criterion compared to
a variety of other algorithms (including a fixed density threshold, a
pure molecular-gas law, a temperature threshold, a Jeans criterion, a
cooling-time criterion and a converging flow criterion).  The star
formation rate follows a volumetric relation:
\begin{equation}
\dot{\rho_*} = \rho_{\rm mol}/t_{\rm ff}
\end{equation}
That is, stars are allowed to form with 100\% \ efficiency per free
fall time.  The star formation is subsequently self-regulated by
stellar feedback, resulting in a time-averaged efficiency on galaxy
scales of $\epsilon_{\rm ff}$ of $\sim 0.005-0.1$\cite{hopkins11b}.

Once stars have formed, they can impact the ISM via various feedback
mechanisms.  Assuming a Kroupa\cite{kroupa02a} stellar initial mass
function, and utilising {\sc starburst99}\cite{leitherer99a} for
luminosity, mass-return and supernova rate calculations as a function
of stellar age and metallicity, we include the following forms of
stellar feedback:
\begin{enumerate}
\item {\it Radiation Momentum Deposition:} At each timestep, the gas near young stars is impacted by a momentum flux given by 
\begin{equation}
\dot{P_{\rm rad}} \approx \left(1-{\rm exp}(-\tau_{\rm UV/optical})\right)(1+\tau_{\rm IR})L_{\rm incident}/c
\end{equation}
Where $\tau_{\rm IR}$ is calculated directly from the simulation, as
$\tau_{\rm IR} = \Sigma_{\rm gas}\kappa_{\rm IR}$, and $\kappa_{\rm
IR}$ = 5(Z/Z$_\odot$) g$^{-1}$ cm$^{2}$.

\item {\it Supernovae and Stellar Winds:} We utilise tabulated Type-1 and Type-II supernovae rates\cite{mannucci06a,leitherer99a}; if a supernova occurs during a timestep, thermal energy and radial momentum are injected within a smoothing length of the star.  Gas and metal return is included as well.  Stellar winds are similarly included with energy, wind momentum, mass and metals deposited within a smoothing length.
\item {\it Photoheating of HII regions:} The production rate of ionising radiation from stars determines the extent of HII regions (allowing for overlapping regions).  These regions are heated to $10^4$ K if the gas is below that threshold.
\end{enumerate}

We utilise the second and third models in the Table to test the
convergence properties of our simulations. One model is run with the
same mass baryonic resolution as our main model (standard resolution;
SR), and one a factor of $\sim 8$ higher resolution (high resolution;
HR).  In ED Fig 7, we show the modeled duty cycle above a given flux
density as a function of flux density for these two models.  We see
that the shortest lived ($\lesssim 200$ Myr) emission spikes present
in the standard resolution model may not be converged in the highest
resolution model.  Notably, emission with longer duty cycles are
either converged, or {\it underpredicted} in our standard resolution
model, suggesting that the relatively long-lived
submillimetre-luminous phase is robust.

We show the $M_*-z$ relation for the central in
ED Fig 8 as compared to
observational constraints\cite{behroozi12a}.  The central galaxy has a
stellar mass a factor $\sim 2$ greater than the observed median
stellar mass for comparable mass halos at this epoch.  The model
galaxy may represent an outlier in the $M_*-z$ relation at these
redshifts.  Indeed, the thickness of the observational constraints
shows the uncertainty, not range in possible values.  Alternatively,
it is possible that the inclusion of feedback from an active galactic
nucleus (AGN) could impact the stellar mass buildup in the galaxy,
though the level to which black hole growth can impact star formation
near the submillimetre-luminous phase is unclear.  Some models have
shown that AGN can grow efficiently in the absence of major
mergers\cite{bellovary13a,anglesalcazar15a}, while other models and
observations suggest that mergers may be necessary to grow massive
holes\cite{hopkins14c,kocevski12a,treister12a}.  The last major merger
before the submillimetre-luminous phase is $\sim 1$ Gyr prior.  Tests
with our low resolution model (m13m14) show that without AGN feedback,
residual star formation drives a factor $\sim 2$ increase in stellar
mass at late times ($z \approx 0-1$).  Finally, we note that a higher
mass resolution model could potentially also result in decreased final
stellar masses.  In our convergence tests, the final $M_*$ (at $z=2$)
of the HR run is $\sim 60\%$ that of the SR run.

\section{Dust Radiative Transfer Calculations}

To calculate the inferred observational properties of our simulated
galaxies, we have developed a dust radiative transfer package, {\sc
powderday}.  In short, {\sc
powderday} takes hydrodynamic simulations of galaxies in
evolution, projects the gas properties onto an adaptive mesh, and
calculates the radiative transfer from the stellar sources through the
dusty interstellar medium until an equilibrium dust temperature is
achieved.

In detail, we identify galaxies utilising {\sc SKID} to locate bound
groups of baryonic particles\cite{governato97a,stadel01a}, and track
their progenitors back in time\cite{thompson15a,thompson15b}.
Galaxies and halos are required to contain at least $64$ particles
each in order to be identified.  We cut out a $200$ kpc (side length)
region around the galaxy of interest, and subdivide the domain into an
adaptive grid with an octree memory structure.  Formally, we begin
with one cell encompassing the entire $8 \times 10^6$ kpc$^3$ radiative transfer
region. The cells then recursively subdivide into octs until there are
a threshold maximum number of gas particles in the cell (we employ
$n_{\rm subdivide,thresh} = 64$, though experiments with $n_{\rm
subdivide, thresh} = 32$ show converged results).  The physical
properties of the gas particles are projected onto the octree using a
spline smoothing kernel\cite{turk11a}.

The spectral energy distribution of stars are calculated on the fly
with the Flexible Stellar Population Synthesis code, {\sc
fsps}\cite{conroy09b,conroy10b} through {\sc python-fsps}, a set of
{\sc python} hooks for {\sc fsps} (\url
{https://github.com/dfm/python-fsps}).  The SEDs are calculated as
simple stellar populations with ages and metallicities determined by
the hydrodynamic simulation, and assuming a Kroupa IMF.

The radiative transfer happens in a Monte Carlo fashion utilising the
three-dimensional dust radiative transfer solver, {\sc
hyperion}\cite{robitaille11a}.  The code uses an iterative methodology
to determine the radiative equilibrium temperature\cite{lucy99a}, and
we determine convergence when the energy absorbed by $99\%$ of the
cells has changed by less than $1\%$ between iterations.  We assume a
dust grain-size distribution comparable to that of the Milky
Way\cite{weingartner01a}, with $R \equiv A_v/E(B-V) = 3.15$.  The dust
emissivities are updated to include an approximation for polycyclic
aromatic hydrocarbons (PAHs) alongside thermal
emission\cite{robitaille12a}.  We assume a constant dust to metals
ratio of $0.4$, motivated by both Milky Way and extragalactic
observational constraints\cite{dwek98a,vladilo98a,watson11a}.

The underlying {\sc hyperion} code has passed the standard benchmarks
for codes of this type\cite{pascucci04a}, and we have found that {\sc
powderday} compares well against other publicly available dust
radiative transfer codes\cite{jonsson06a,jonsson10a,torrey15a} in test
starburst SPH galaxy merger simulations.

\section{Parameter Choices}

In ED Fig 9, we present a
number of tests of our parameter choices for the radiative transfer
calculations.  We show the predicted $850 \mu$m light curve from our
lowest resolution model (m13m14) utilising both fiducial parameters,
as well as three parameter choice variations.

We first ask whether our chosen radiative transfer grid size affects our
principle results.  Our fiducial model is a $200$ kpc (on a side) box
cut out of the global cosmological simulation centred on the halo of
interest.  This size was chosen to reflect a rough average of the
typical (sub)millimetre beam sizes typically used to detect SMGs.  For
example, assuming Planck 2013 cosmological parameters\cite{planck14a},
the Submillimetre Common-Use Bolometer Array (SCUBA) on the James
Clerk Maxwell Telescope (JCMT) has a $15^"$ full width at half maximum
(FWHM) beam at $850 \mu$m.  At $z=2$ this corresponds to $\sim 128$
kpc.  At the same redshift, the beam of AzTEC and LABOCA at 1 mm on
the JCMT corresponds to $\sim 163$ kpc ($19^{``}$); The South Pole
Telescope (SPT) has a beamsize of $540$ kpc at 1.4 mm ($63^{``}$); and
Herschel's SPIRE instrument ranges from $154-308$ kpc ($250-500 \ \mu$m;
$18-36^{``}$).

Because a few notable beam sizes (of particular relevance, the SCUBA
beam) are smaller than our assumed box size of $200$ kpc, we have run
an additional model with box length $100$ kpc (all other parameters
exactly the same).  We highlight the resultant 850 $\mu$m light curve
from this model in the top right panel of ED
Fig 9.  When comparing to our fiducial model, it
is apparent that our results are robust to the highest resolution
beams that have been used for SMG surveys at single dish facilities to
date.

We additionally investigate whether our inclusion of PAHs in our model
makes any difference to the calculated submillimetre-wave flux density
of our model galaxy.  This is presented in the bottom left panel of
ED Fig 9.  Again, we note minimal impact on the submillimetre
SED of our model.

Finally, we ensure that our results are converged with the number of
photons emitted.  We fiducially run $10^7$ photons per grid (roughly
100 per cell).  In the bottom-right, we show the results from a run
with $10^8$ photons per grid, and show that the results are robust
against this parameter choice.

\section{Relation to other Models}

Historically, the methods used, and physical models for SMG formation
in numerical simulations are quite varied. Here, we summarise these
methods and results, and place our own model into this context.
Broadly, there are three classes of SMG formation models: cosmological
semi-analytic models (SAMs), idealised non-cosmological simulations,
and cosmological hydrodynamic models.  The present model falls into
the latter category.  Our presented model is the first
self-consistent cosmological simulation with baryons and bona fide
radiative transfer to form a submillimetre galaxy with physical
properties comparable to those observed.

The initial forays into this field were typically with SAMs.  This
owes to the fact that SAMs are computationally inexpensive, and allow
for a large search in physical parameter space relatively easily. SAMs
either utilise analytic halo merger trees, or directly simulate them,
and then employ analytic prescriptions to describe the central
galaxies.  The Durham SAM\cite{baugh05a,gonzalez11a} couples galaxies
formed in a semi-analytic model with dust radiative transfer.  These
simulations model galaxies that have axisymmetric geometries that
consist of a disc and a bulge.  Young stellar populations are assumed
to still be enveloped in their birth clouds, and thus experience
additional attenuation.  This model suggests that SMGs typically owe
to $\sim 22\%$ major mergers, the remainder as minor mergers, and that
the stellar initial mass function is flat during the starburst.  The
typical duty cycle for the submillimetre-luminous phase is $\sim 100$
Myr (a factor $\sim 7.5$ lower than found in our work), galaxies extremely
gas rich ($f_{\rm gas} \sim 75\%$), and stellar masses a factor $\sim
10$ lower than our model ($M_* \sim 2. \times 10^{10} M_\odot$).
While the stellar masses of SMGs are
debated\cite{hainline11a,michalowski12a,michalowski14a}, the gas
fractions appear to be uniformly lower in
observations\cite{bothwell12a,narayanan12c,fu13a,ivison13a,tacconi13a},
and a flat stellar initial mass function likely ruled out by CO
dynamical mass measurements\cite{tacconi08a}.

As an alternative to cosmological simulations, a number of works have
explored SMG formation in idealised
simulations\cite{chakrabarti08a,narayanan09a,narayanan10a,hayward11a,hayward13a}. These
works evolve hydrodynamic models of idealised discs and mergers over a
range of merger mass ratios, and combine these with dust radiative
transfer simulations\cite{jonsson10a}.  These models infer halo masses
and stellar masses for SMGs comparable to those modeled here.  This
said, in the idealised galaxy models, $\sim 30-70\%$ of the SMGs (flux
dependent) owe to merger-driven starbursts, substantially higher than
what is found for our model.  Some works\cite{hayward13a} have noted
that binary mergers that cause SMGs may break up into multiples at
high-resolution owing to the contribution to the total flux of
individual inspiralling discs.  Because idealised simulations are
non-cosmological in nature, comparing the multiplicity inferred from
these to our models is difficult: the major merger multiplicity can
only be 2 when considering galaxies at the same redshift.  On the
other hand ED Fig 2 suggests potentially
larger multiplicity can be observed for physically associated clumps.

To fully capture the cosmic environment of SMGs in formation, as well
as their baryonic structure and morphology, cosmological hydrodynamic
simulations are likely the best tool.  Thusfar, cosmological
hydrodynamic simulations used to simulate SMGs have not employed
direct radiative transfer models\cite{dave10a}.  As such, inferring
when a galaxy is an SMG in cosmological simulations has necessitated
the use of parameterised emission models, such as assumed grey-body
emission laws\cite{shimizu12a}, or star formation rate
thresholds\cite{dave10a}.  The physical properties for SMGs derived
from the most extensive of these works\cite{dave10a} (i.e. $M_*$,
$M_{\rm DM}$, and $f_{\rm gas}$) are similar to the model presented
here, though with roughly a factor $\sim 3$ difference in SFR.

\section{Code Availability}
We have made {\sc powderday} available
at \url{https://bitbucket.org/desika/powderday}, and {\sc GIZMO}
available at \url{https://bitbucket.org/phopkins/gizmo}

\clearpage
\setcounter{page}{1}
\setcounter{figure}{0}
\setcounter{table}{0}
\captionsetup[figure]{labelformat=empty}% redefines the caption setup of the figures environment in the beamer class.

\renewcommand{\thefigure}{Extended Data \arabic{figure}}
\renewcommand{\thetable}{Extended Data \arabic{table}}

%\begin{center}
%{\bf \Large \uppercase{Extended Data} }
%\end{center}

\begin{table*}
\footnotesize
\caption{ \textbf{Summary of Model Galaxies.}  $M_*$ and $M_{\rm halo}$ refer to the stellar and halo mass at $z=2$; $\epsilon_b$ and $\epsilon_{\rm DM}$ refer to the minimum force softening lengths for baryons and dark matter particles.}
\label{table:models}
\begin{tabular}{|l|l|l|l|l|l|l|l|l|}
\hline\hline
{\bf Model Name} & {\bf Model Purpose} & {\bf $M_*$ (z=2)}  & {\bf $M_{\rm halo}$ (z=2)} &  {\bf $m_{\rm b}$} & {\bf $m_{\rm DM}$} & {\bf $\epsilon_{\rm b}$} & {\bf $\epsilon_{\rm DM}$}& {\bf Final Redshift} \\ 
               &                     & $M_\odot$           & $M_\odot$                 &  $M_\odot$       &  $M_\odot$     & pc                    & pc                     &\\
\hline
\hline
B100    & Main Model & $4 \times 10^{11}$ & $3 \times 10^{13}$ & $2.7\times10^5$&$1.3\times10^6$ &9 & 142 &  2 \\

TL37 SR       & Resolution Test & $8 \times 10^{10}$ & $7 \times 10^{12}$ & $2.7 \times 10^5$ &$1.3 \times 10^6$ & 9 & 142 & 2 \\
TL37 HR       & Resolution Test & $5 \times 10^{10}$ & $7 \times 10^{12}$ & $3.3 \times 10^4$ &$1.7 \times 10^5$ & 9 & 142 & 2 \\
m13m14     & RT Parameter Survey & $3 \times 10^{11}$ & $7 \times 10^{12}$ & $4.4 \times 10^6$ & $2.3 \times 10^7$ & 70 & 700 & 0.2 \\
\hline\hline
\end{tabular}
\end{table*}

\begin{figure*}
\centerline{
\includegraphics[width=0.9\textwidth]{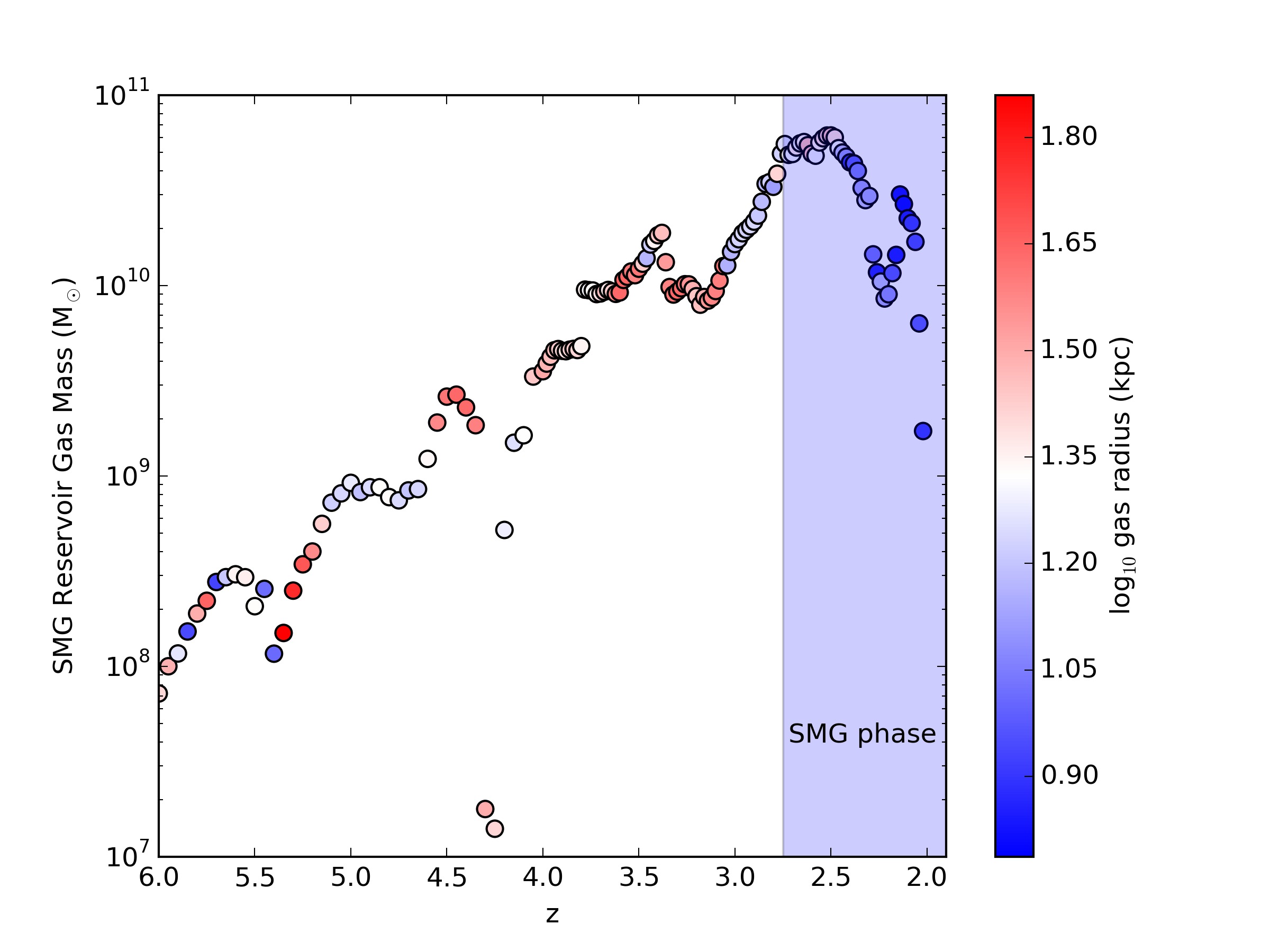}
}
\vspace{-4mm}
\caption{\textbf{Extended Data Figure 1: Mass of gas in central galaxy that will be consumed during SMG starburst as a function of $z$.}  The gas mass consumed during the starburst is calculated by tracking the evolution of gas particles that turn into stars during the SMG phase ($z \approx 2-2.7$), and is only measured for the central galaxy itself (i.e. gas ejected into the halo is not included).   The colours denote the median scale height from the galaxy centre of mass.   The SMG gas reservoir follows a cycle of being pushed outward followed by re-accretion.    \label{figure:ed_accretion} }
\vspace{-4mm}
\end{figure*}

\begin{figure*}
\centerline{
\includegraphics[width=0.9\textwidth]{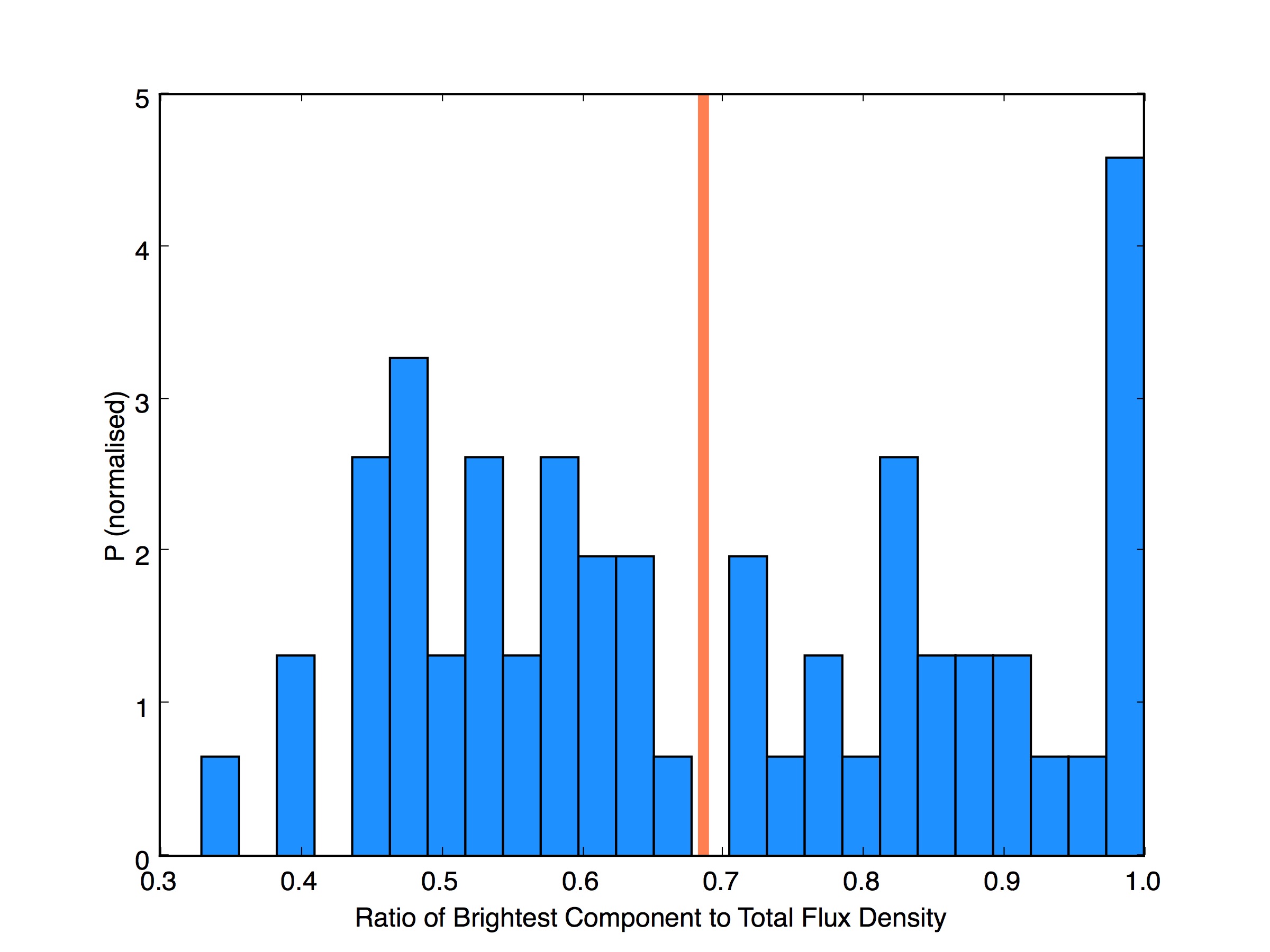}
}
\vspace{-4mm}
\caption{\textbf{Extended Data Figure 2: Predicted contribution of submillimetre-luminous region components to total flux density.}  Submillimetre-luminous regions often break up into multiples.  Shown is a histogram of the ratio of the brightest component to the total flux density from the region, with the average denoted with the vertical line.  The region is generally dominated by one component, though smaller subhalos can contribute on average $\sim 30\%$ of the observed flux density.  \label{figure:multiplicity} }
\vspace{-4mm}
\end{figure*}

\begin{figure*}
\centerline{
\includegraphics[width=0.9\textwidth]{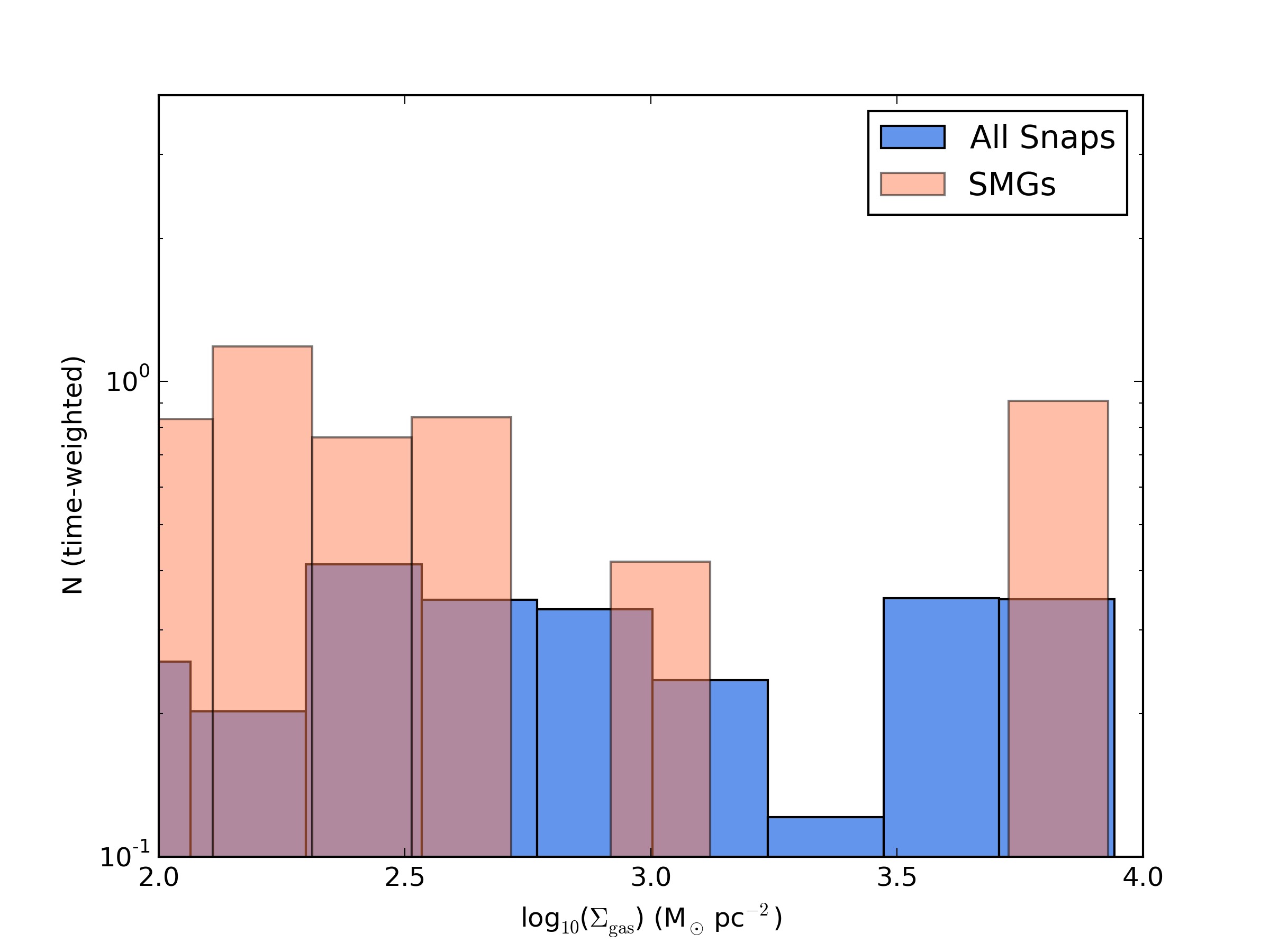}
}
\vspace{-4mm}
\caption{\textbf{Extended Data Figure 3: Gas surface density for central submillimetre galaxy.}  The blue histograms show the time-weighted distribution of surface densities during all phases, whilst the orange show the same for the submillimetre-luminous phase.    We predict that the submillimetre-luminous phases do not have dramatically different surface density distributions compared to the non-submillimetre-luminous phases.  This feature may be tentatively observed\cite{narayanan12a,casey14a}. \label{figure:ed_gas_surface_density} }
\vspace{-4mm}
\end{figure*}

\begin{figure*}
\centerline{
\includegraphics[width=0.9\textwidth]{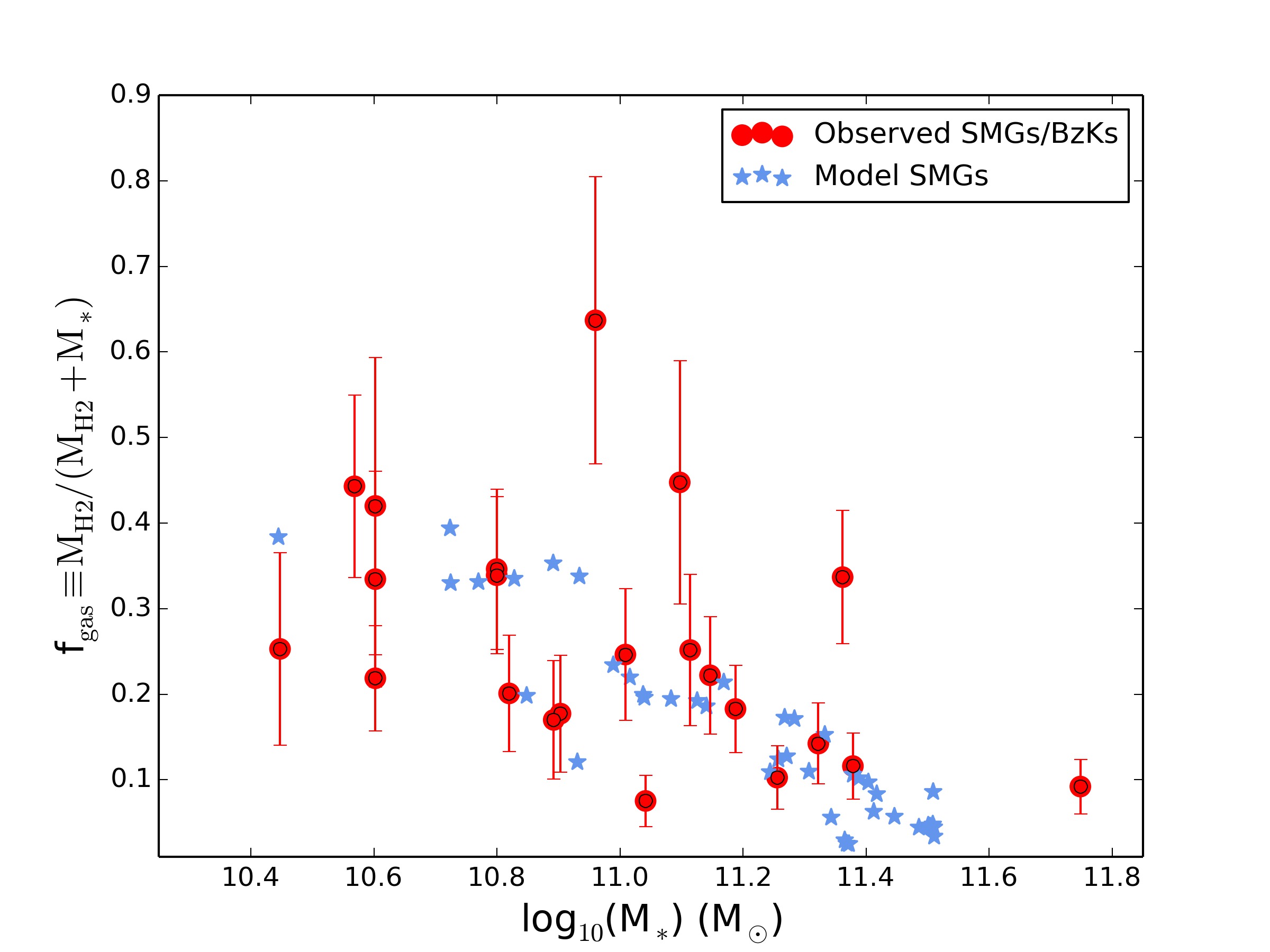}
}
\vspace{-4mm}
\caption{\textbf{Extended Data Figure 4: Molecular gas fraction as a function of galaxy stellar mass.}  Blue stars show individual snapshots of central submillimetre galaxy whilst red circles with error bars ($1\sigma$) show observations with direct CO (J=1-0) measurements (to avoid complications in converting from higher-lying CO rotational lines to the ground state for a mass conversion).    Both observations and our model show a declining molecular gas fraction with increasing galaxy mass, with a typical range of $f_{\rm gas} = 0.1-0.4$ for galaxies of SMG mass.\label{figure:fgas}}
\vspace{-4mm}
\end{figure*}

\begin{figure*}
\centerline{
\includegraphics[width=0.9\textwidth]{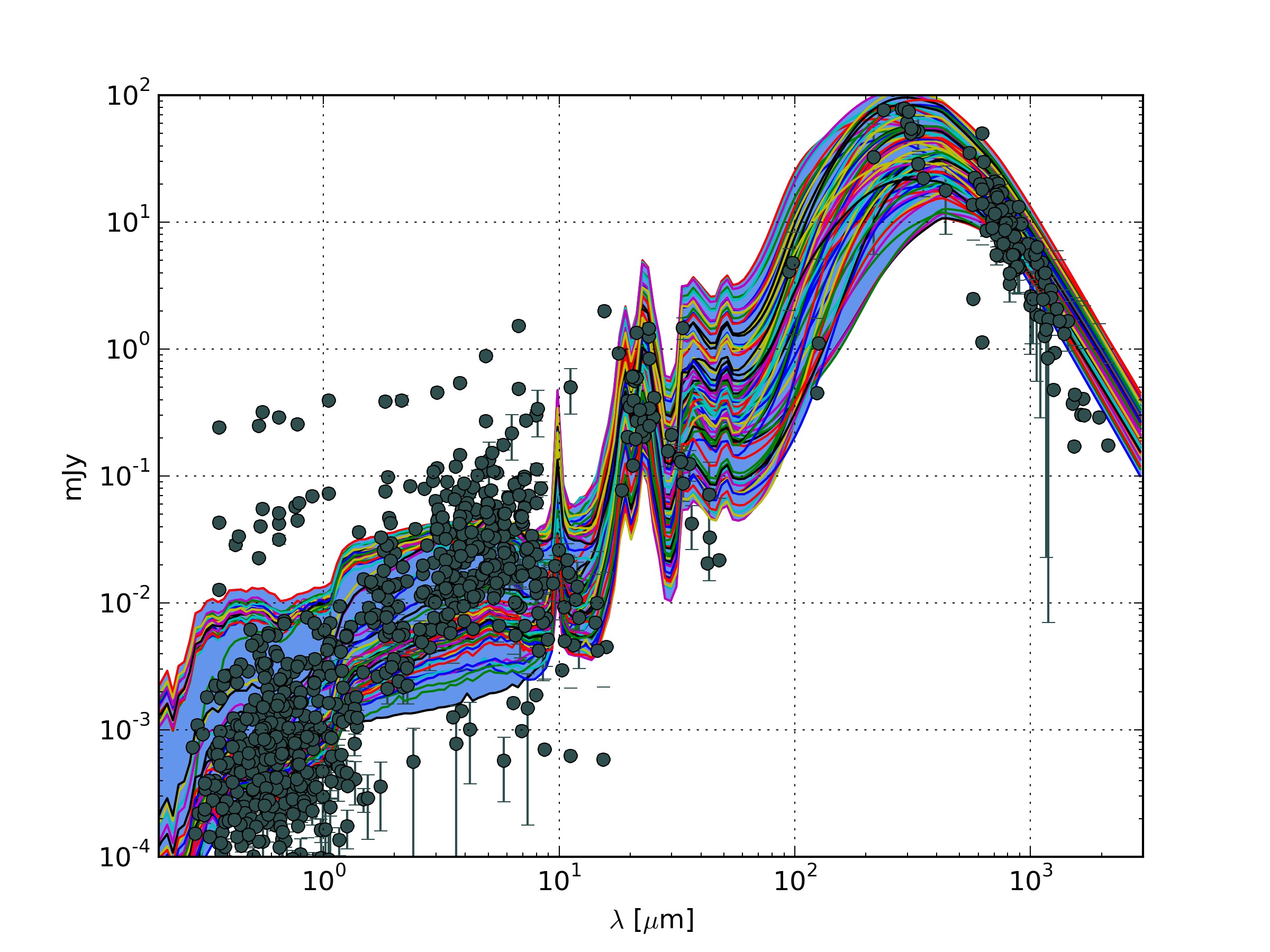}
}
\vspace{-4mm}
\caption{\textbf{Extended Data Figure 5: Predicted spectral energy distribution (SED) for central submillimetre galaxy.}  The blue shaded region shows the range of SEDs for all snapshots that satisfy the fiducial $F_{\rm 850 \mu m} > 5$ mJy submillimetre galaxy selection criteria, whilst the grey points with error bars ($1\sigma$) are a compilation of observed data.  The individual coloured lines show the SEDs for individual submillimetre luminous snapshots.  The data and models are redshifted to a common redshift $z=2$.  The model and data compare well, and the model suggests a diverse range of SMG SEDs.\label{figure:sed}}
\vspace{-4mm}
\end{figure*}

\begin{figure*}
\centerline{
\includegraphics[width=0.9\textwidth]{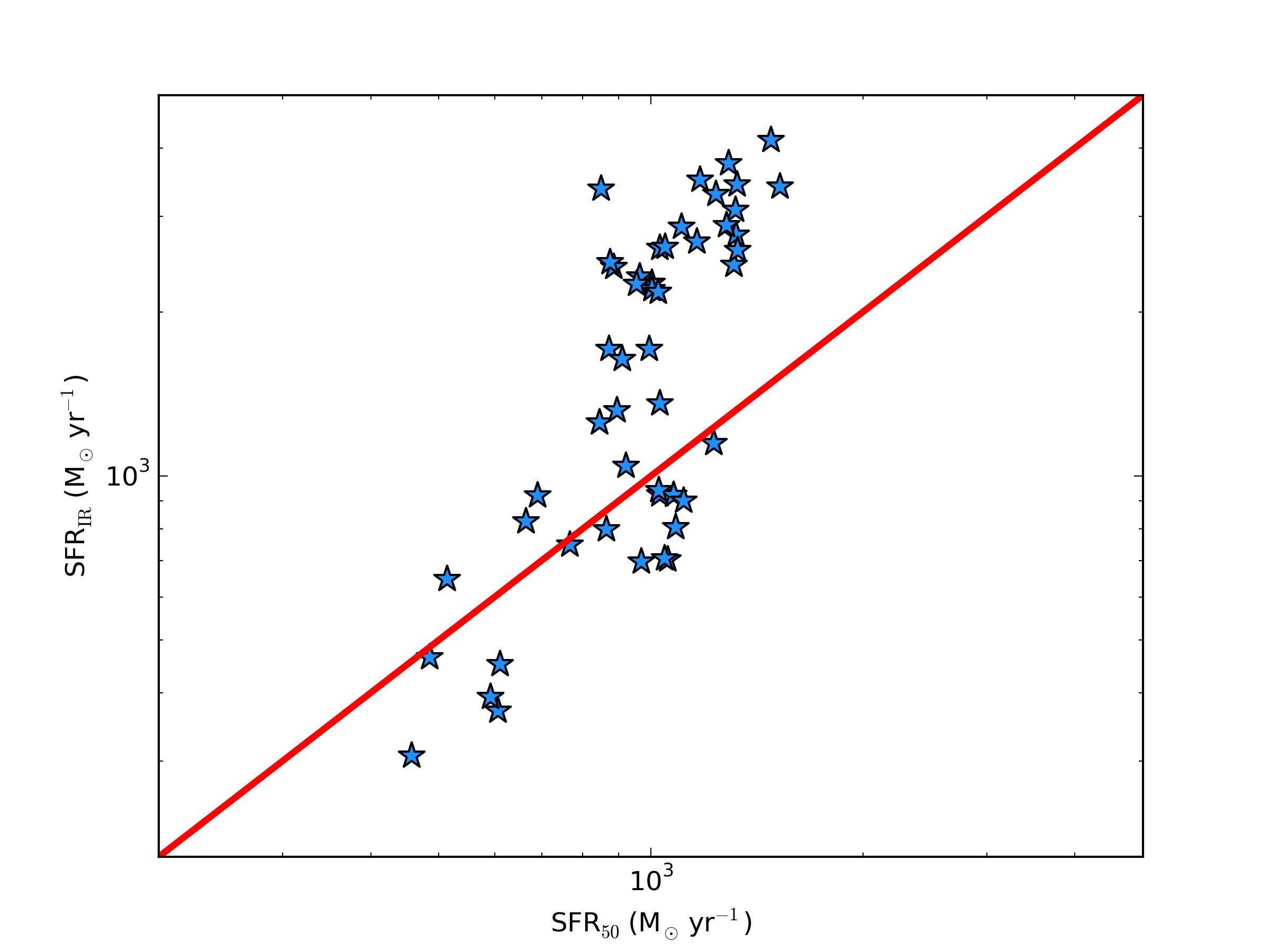}
}
\vspace{-4mm}
\caption{\textbf{Extended Data Figure 6: Overestimate of SFR of High-$z$ SMGs}. The ordinate denotes the SFR as determined from the infrared SED\cite{kennicutt12a}, while the abscissa shows the SFR averaged over the last $50 $Myr in the simulations.   Up to an SFR of $\sim 800 M_\odot $yr$^{-1}$ the two correspond well.  At
  higher SFRs, however, there is a dramatic departure owing to
  substantial contribution to the infrared luminosity by older
  stars. \label{figure:sfr_ir}}
\vspace{-4mm}
\end{figure*}

\begin{figure*}
\centerline{
\includegraphics[width=0.9\textwidth]{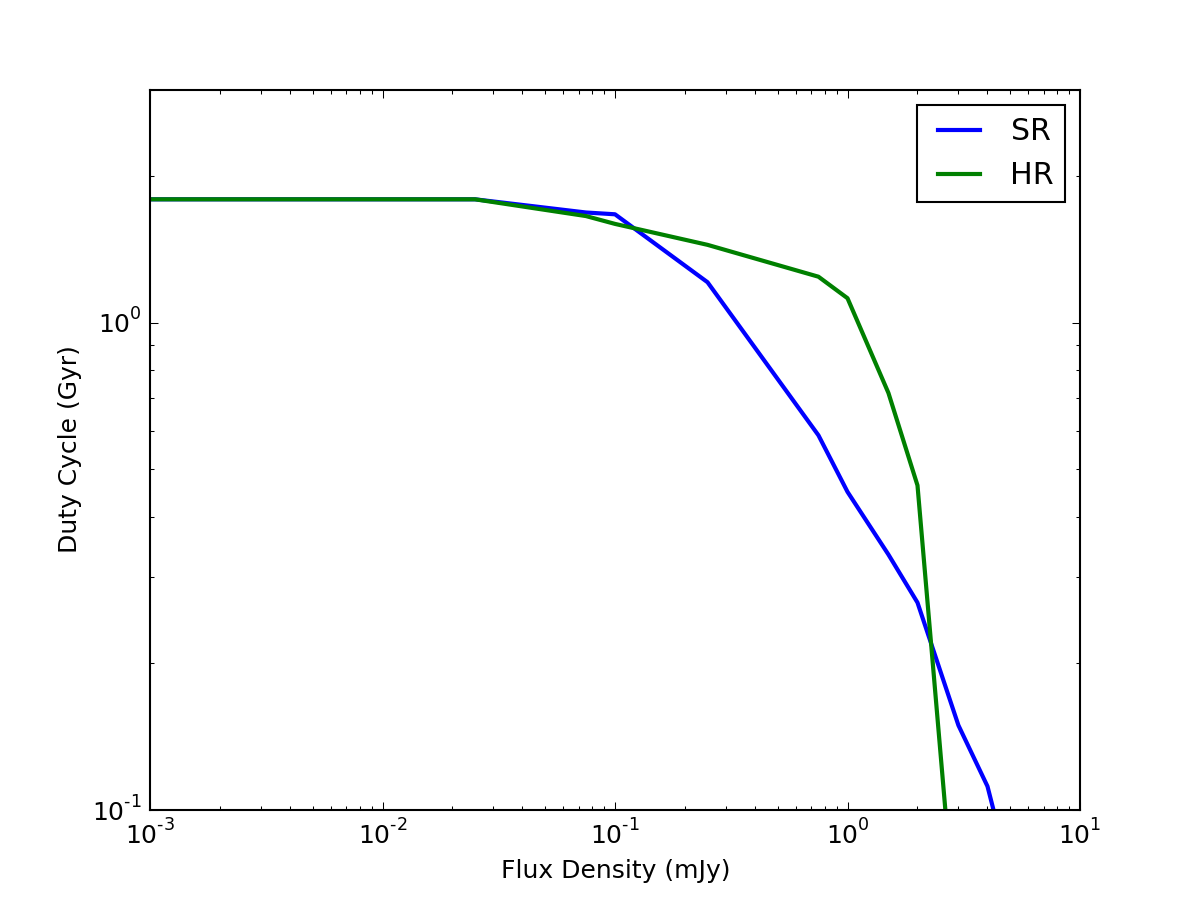}
}
\vspace{-4mm}
\caption{\textbf{Extended Data Figure 7: Resolution tests for hydrodynamic zoom simulations.}  Lines show $850 \mu$m duty cycle above a given flux density as a function of flux density for our resolution test models presented in the Methods section.   SR denotes our standard resolution (the resolution of our main model) whilst HR is one level higher refinement.\label{figure:convergence} }
\vspace{-4mm}
\end{figure*}

\begin{figure*}
\centerline{
\includegraphics[width=0.9\textwidth]{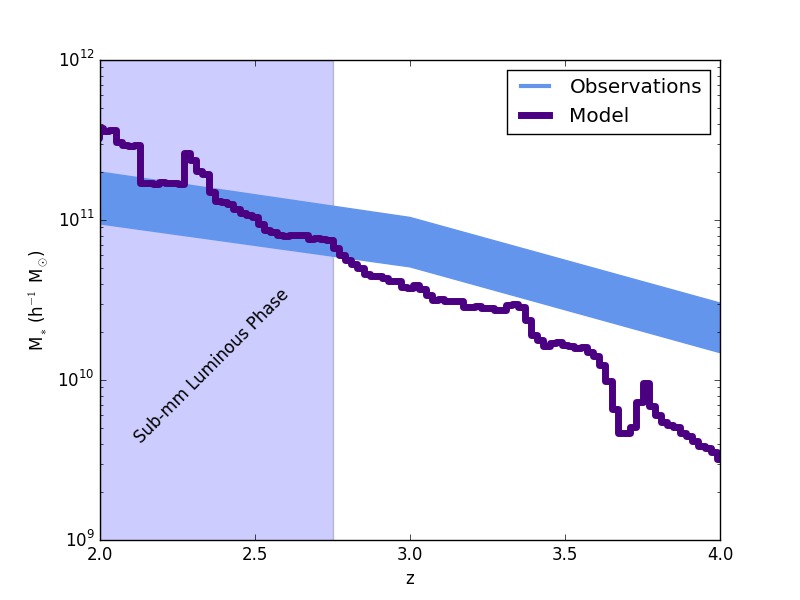}
}
\vspace{-4mm}
\caption{\textbf{Extended Data Figure 8: Stellar Mass - Redshift Relation for Model Galaxy.}
Purple line shows model whilst blue filled region shows observational
  constraints from an abundance matching
  assumption\cite{behroozi12a}. The model and observations are in
  reasonable agreement, especially during the submillimetre-luminous
  phase (purple shaded region).  At late times, the stellar mass of
  the galaxy is a factor $\sim 2$ higher than the median observed
  galaxy.\label{figure:mstar_mhalo} }
\vspace{-4mm}
\end{figure*}

\begin{figure*}
\centerline{
\includegraphics[width=0.9\textwidth]{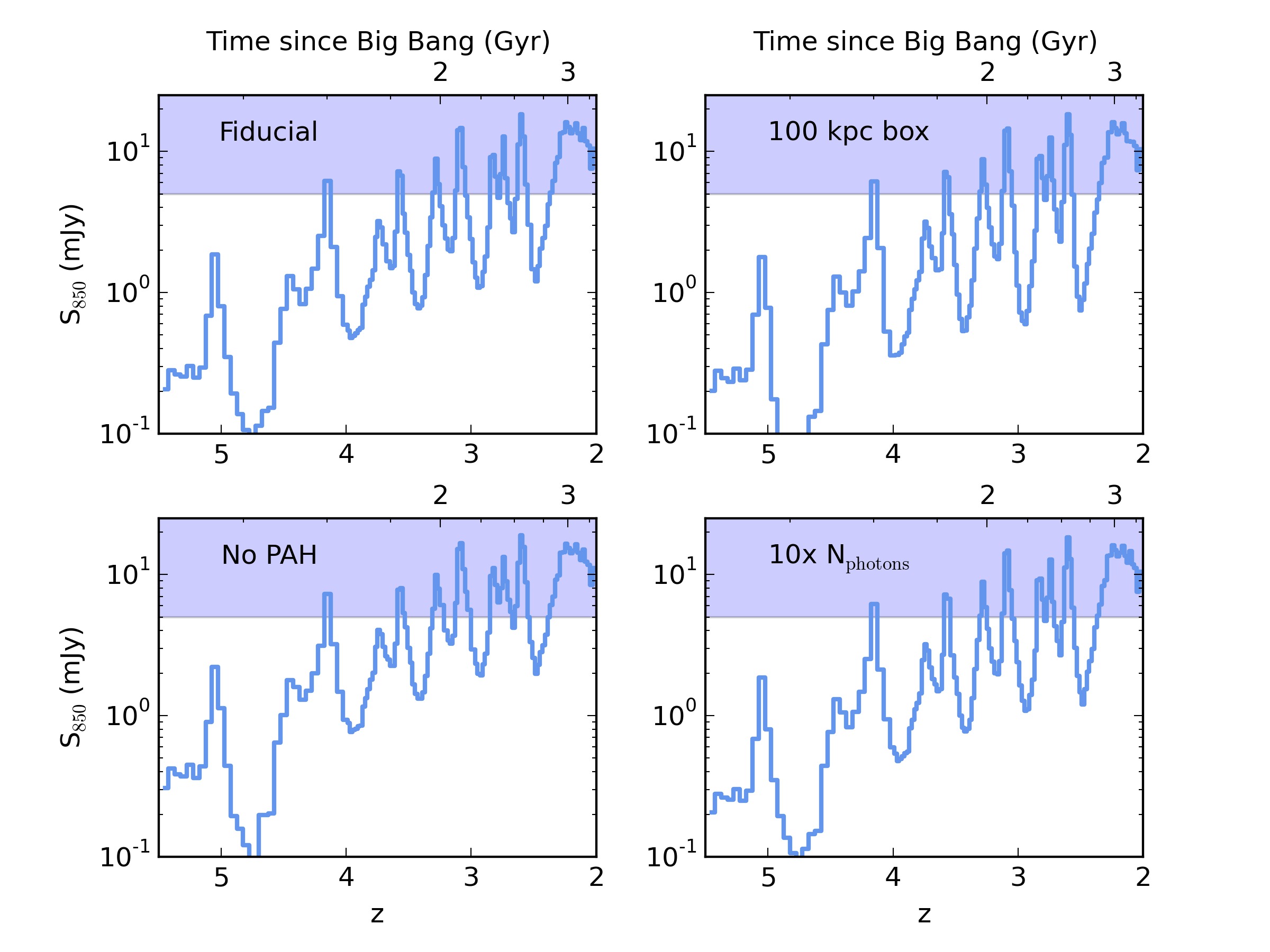}
}
\vspace{-4mm}
\caption{\textbf{Extended Data Figure 9: Tests of parameter choices for radiative transfer
    calculations.}  The simulated galaxy for these tests is our lowest
  resolution cosmological simulation (m13m14).  Each panel shows the
  850 $\mu$m flux density lightcurve of the tested model, with time
  noted on the abscissa (redshift on the bottom, time since the Big
  Bang on top).  In all panels, the purple shaded region denotes
  $S_{\rm 850} \geq 5$ mJy, the canonical selection criteria for SMGs.
  {\it Top Left:} Our fiducial set of parameters; {\it Top Right:} Simulation with
  a 100 kpc (on a side) emission region instead of 200 kpc; {\it Bottom Left:}
  Simulation with our model for PAHs turned off; {\it Bottom Right:}
  Fiducial simulation run with ten times the number of
  photons.\label{figure:sfr_850_si}}
\vspace{-4mm}
\end{figure*}

\end{document}